\documentclass[preprint,superscriptaddress,showkeywords,nofootinbib]{revtex4}

\usepackage{amssymb,amsmath,graphicx,dcolumn,bm,hyperref}
\usepackage{CJK}
\usepackage{color}

\begin{document}

%\begin{CJK*}{GBK}{song}

\newcommand*{\pku}{School of Physics and State Key Laboratory
of Nuclear Physics and Technology, \\Peking University, Beijing
100871}\affiliation{\pku}
\newcommand*{\CHEP}{Center for High Energy
Physics, Peking University, Beijing 100871}\affiliation{\CHEP}

\title{A Proof of First Digit Law from Laplace Transform\footnote{Published in Chinese Physics Letters 36 (2019) 070201 }\footnote{Supported by the National Natural Science Foundation of China under Grant No.~11475006}}

\author{Mingshu Cong}\affiliation{\pku}
\author{Bo-Qiang Ma}\email[Corresponding author. Electronic address:~]{mabq@pku.edu.cn}\affiliation{\pku}\affiliation{\CHEP}
%×÷ÕßÐÅϢΪ   Mingshu Cong (´ÔÃ÷Êæ)£¬ Bo-Qiang Ma (Âí²®Ç¿)
%×÷ÕßÁªÏµ·½·¨£º mabq@pku.edu.cn; mabqpku@163.com, Tel: 62765708, 13671336280

\begin{abstract}

The first digit law, also known as Benford's law or the significant digit law, is an empirical
phenomenon
that the leading digit of numbers from real world sources favors small ones in a form
% over bigger ones with a pattern as
%that the occurrence of digits 1 to 9 as the leading
%digit of numbers from real world sources is often not uniformly
%distributed, but instead, distributed unevenly according to a logarithmic form
$\log(1+{1}/{d})$, where $d=1, 2, ..., 9$.  Such a law keeps elusive for over one hundred years
because it was obscure whether this law is due to the logical consequence of the number system or
some mysterious mechanism of the nature. We provide a simple and elegant proof of this law from the application of the Laplace
transform, which is an important tool of mathematical methods in physics. We reveal that the first digit law
is originated from the basic property of the number system, thus it should be attributed as a basic mathematical knowledge for wide applications.

\end{abstract}

%\pacs{02.50.Cw, 05.20.-y}

%\vspace{20mm} Keywords: first digit, Benford's law, mathematical proof, Laplace transform

\maketitle

%\end{CJK*}

\begin{figure}[h]
\begin{center}
\includegraphics[width=10cm]{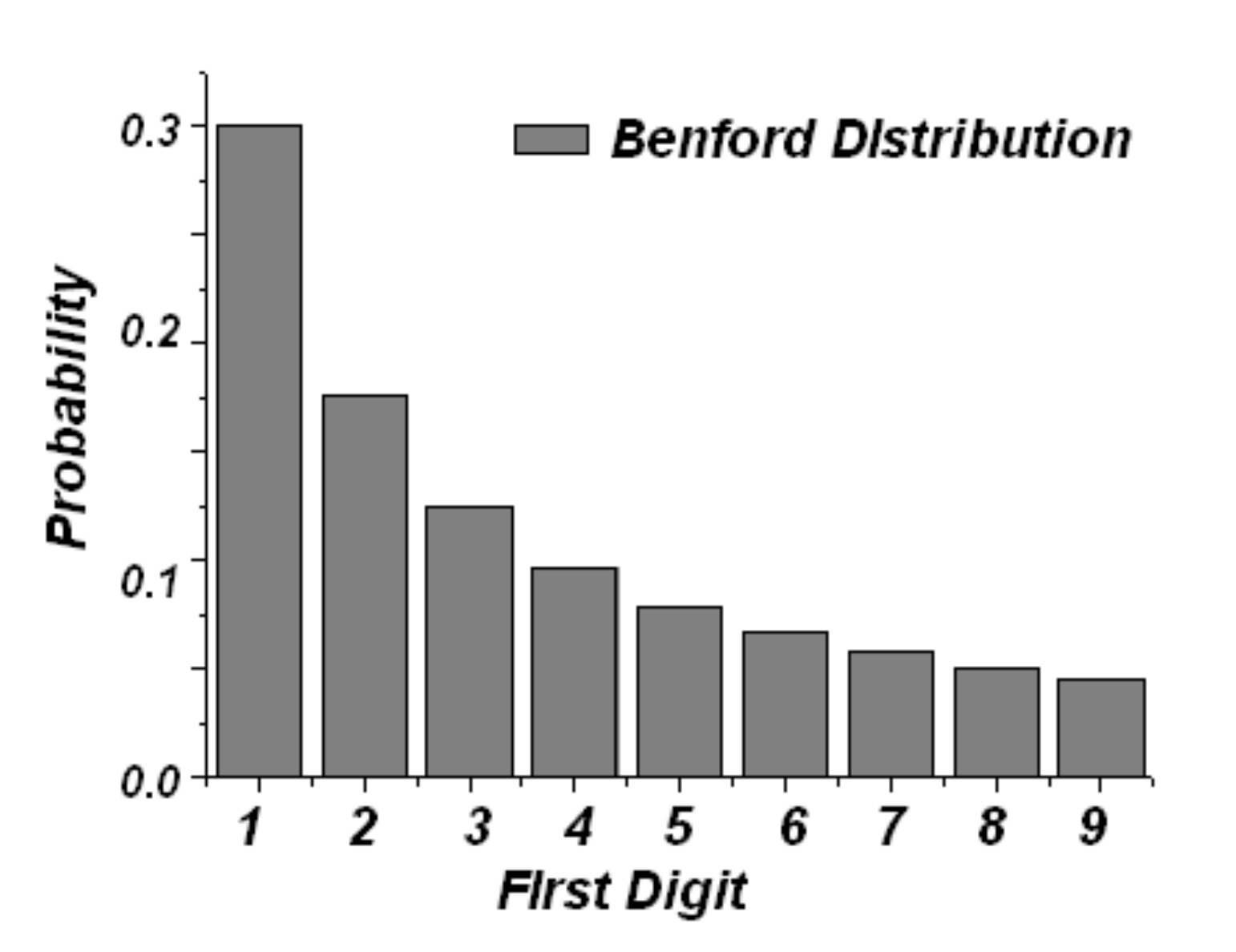}
\caption{Benford's law of the first digit distribution, from which
we see that the probability of finding numbers with leading digit 1
is larger than that with $2, ..., 9$ respectively.}\label{Benfordfigure}
\end{center}
\end{figure}

%\section*{Introduction}
The first digit law, which is also called the significant
digit law or Benford's law, was first noticed by Newcomb in 1881~\cite{n81}, and then
re-discovered independently by Benford in 1938~\cite{b38}.
It is an empirical observation that the first digits of natural
numbers prefer small ones rather than a uniform distribution as
might be expected. More accurately, the probability that a number
begins with digit $d$, where $d=1,2,...,9$ respectively, can be
expressed as
\begin{equation}
\label{benford}
P_d = \log(1+\frac{1}{d}),\, \, d=1, 2, ..., 9,
\end{equation}
as shown in Fig.~\ref{Benfordfigure}.

Empirically, the populations of countries,
the areas of lakes, the lengths of rivers, the arabic numbers on the
front page of a newspaper~\cite{b38}, physical
constants~\cite{bk91}, the stock market indices~\cite{l96}, file
sizes in a personal computer~\cite{tfgs07},
%survival distributions~\cite{lse00},
%{\it et al.},
etc.,
all conform to the
peculiar law well.
%Due to the powerful data analyzing tools provided by computer science, %Surprisingly,
Benford's law has been verified to hold true for a vast number of examples in
various domains, such as economics~\cite{l96}, social
science~\cite{lse00}, environmental science~\cite{b05},
biology~\cite{biology}, geology~\cite{geology},
astronomy~\cite{sm10a}, statistical physics~\cite{sm10b,sm10c},
nuclear physics~\cite{bmp93ejp,nr08,ren09,liu2011benford,hui2011benford}, particle physics~\cite{sm09a},
and some dynamical systems~\cite{tbl00,bbh05,b05dcds}. Also, there have
been many explorations on applications of the law in various
fields, mainly to detect data and judge their reasonableness,
such as in distinguishing and ascertaining fraud in taxing and
accounting~\cite{n96,n99,rr03}
%fabrication in clinical trials~\cite{marzouki2005},
%authenticity of the pollutant concentrations in ambient air~\cite{b05},
%electoral cheats or voting anomalies~\cite{tfgs07},
and falsified data in scientific
experiments~\cite{d07a}.
%Moreover, the first digit law is applied in
%computer science for speeding up calculation~\cite{barlow1985},
%minimizing expected storage space~\cite{s88,bh07}, analyzing the
%behavior of floating-point arithmetic algorithms~\cite{bh07}, and
%especially for various studies in the image domain~\cite{j01,f07}.
%and especially on earthquake prediction~\cite{geophy}.
%Recently, Benford's law is also applied to earthquake
%prediction~\cite{geophy}.

%Theoretically,
Benford's law has several elegant properties. It
is scale-invariant~\cite{p61,bhm08}, which means that the law does
not depend on any particular choice of units. This law is also
base-invariant~\cite{h95a,h95b,h95c}, which means that it is
independent of the base $b$ %.
with a general form
%In a similar form to that in the
%decimal system ($b$=10), in the binary system ($b$=2), octal
%system ($b$=8), or other base system, the data all fit the general Benford's law,
\begin{equation}\label{benbase}
P_d = \log_{b}(1+\frac{1}{d}), \,\, d=1, 2, ..., {b-1}, \,\, \mathrm{for}\,\, b \ge 2.
\end{equation}
%It has been proved that ``scale-invariance implies
%base-invariance''~\cite{p61,h95a} and ``base-invariance implies
%Benford's law''~\cite{h95b} mathematically in the framework of
%probability theory.
The law is also found to be power-invariant~\cite{sm09a}, i.e., any
power ($\neq 0$) on numbers in the data set does not change the
first digit distribution.
Though there have been many studies on Benford's law~\cite{Benfordsite},
% with numerous breakthroughs~\cite{Benfordsite},
the underlying reason for the
success of this law remains elusive for more than one hundred
years.
%A general idea in this field is that Benford's law will be
%satisfied under some hypotheses. Driven by this belief, Engel and
%Leuenberger proved that exponentially distributed numbers obey this
%law within bounds of 0.03 ~\cite{el03}. This fact has directive
%significance to the method in this paper.
It was unclear whether Benford's Law is due to
some unknown mechanism of the nature or it is merely a logical
consequence of human number system.

However, the situation has been changed
due to the appearance of a general derivation of Benford's law from the application
of the Laplace transform~\cite{Cong19PLA}, where a strict version of Benford's law
is derived as composed of a Benford term and an err term. The Benford term explains
the prevalence of Benford's law and the err term leads to derivations from the law with four
categories of number sets. It is the purpose of this Letter to provide a more simple and elegant
version of the derivation of Benford's law compared to Ref.~\cite{Cong19PLA}. Through this derivation,
it is easier to understand the rationality of Benford's law.
% from the Laplace transform, which is an important tool of mathematical methods of physics.
We reveal that the first digit law can be derived as the main term from the Laplace transform.
This explains why Benford's law is so successful for many number sets.
We perform similar analysis on the
regularities of the second digit and $i$th-significant digit
distributions, and extend the law to a more general rule for the
first several digit distribution. We also estimate the error term and point out conditions for
the validity of this law.

%A fundamental question is whether Benford's Law is due to
%some unknown mechanism of the nature or it is only a logical
%consequence of digital system.
%The purpose of this paper is to assert that the latter case
%is the right answer for Benford's law.

%We provide an intuitive
% mathematical explanation of
%Benford's law with the application of Laplace transform. We
%reveal that the first digit law is only a logical consequence from the basic
%property of digital system, instead of specific quality of a common
%distribution or any unknown mechanism.
% Then we perform similar analysis on the
%regularities of the second digit and $i$th-significant digit
%distributions, and extend the law to a more general rule for the
%first several digit distribution. Finally, we estimate the error for
%our deviation and point out some conditions for
%the validity of this law.

%\section*{An Intuitive Explanation of the First Digit Law}\label{mantissa}

For simplification, we constrain ourselves to the decimal system first.
Let $F(x)$ be an arbitrary normalized probability density
function defined on positive real number set $\mathbb{R}^+$. (Here we use the capital letter
F instead of the lowercase one, as opposed to the convention.) Of
course, in the real case, the variable $x$ may be negative or
bounded, but this is not harmful to our derivation. When $x$ can be negative,
 we can use the probability density function of its absolute value, keeping results
 unchanged.

In the decimal system, the probability $P_d$ of finding a number
whose first digit is $d$ is the sum of the probability that it is
contained in the interval $[d \cdot 10^n, (d+1) \cdot 10^n)$ for all
integer $n$, therefore $P_d$ can be expressed as
\begin{equation}\label{sum}
P_d = \sum_{n = -\infty}^{\infty} \int_{d \cdot 10^n}^{(d+1) \cdot
10^n} F(x) {\rm d} x \,,
\end{equation}
which can also be rewritten as
\begin{equation}\label{sumDisForm}
P_d = \int_{0}^{\infty} F(x)g_d(x) {\rm d} x \,,
\end{equation}
with $g_d(x)$ being a new density function whose significance will
be clear in the following. (Here the lowercase letter is used, due to conventions for Laplace transform in the following sections.) Adopting the Heaviside step function,
\begin{eqnarray}
\eta(x)= \left\{
\begin{array}{ll}
    1 ,  &   \mbox{if $x \ge 0$, } \\
    0  ,   &  \mbox{if $x<0$},
\end{array}
\right.
\end{eqnarray}
we can write $g_d(x)$ as
\begin{equation}\label{gDefine}
g_d(x)=\sum_{n = -\infty}^{\infty}[\eta (x-d \cdot 10^n)- \eta
(x-(d+1) \cdot 10^n)] \,.
\end{equation}

\begin{figure}[h]
\begin{center}
\includegraphics[width=10cm]{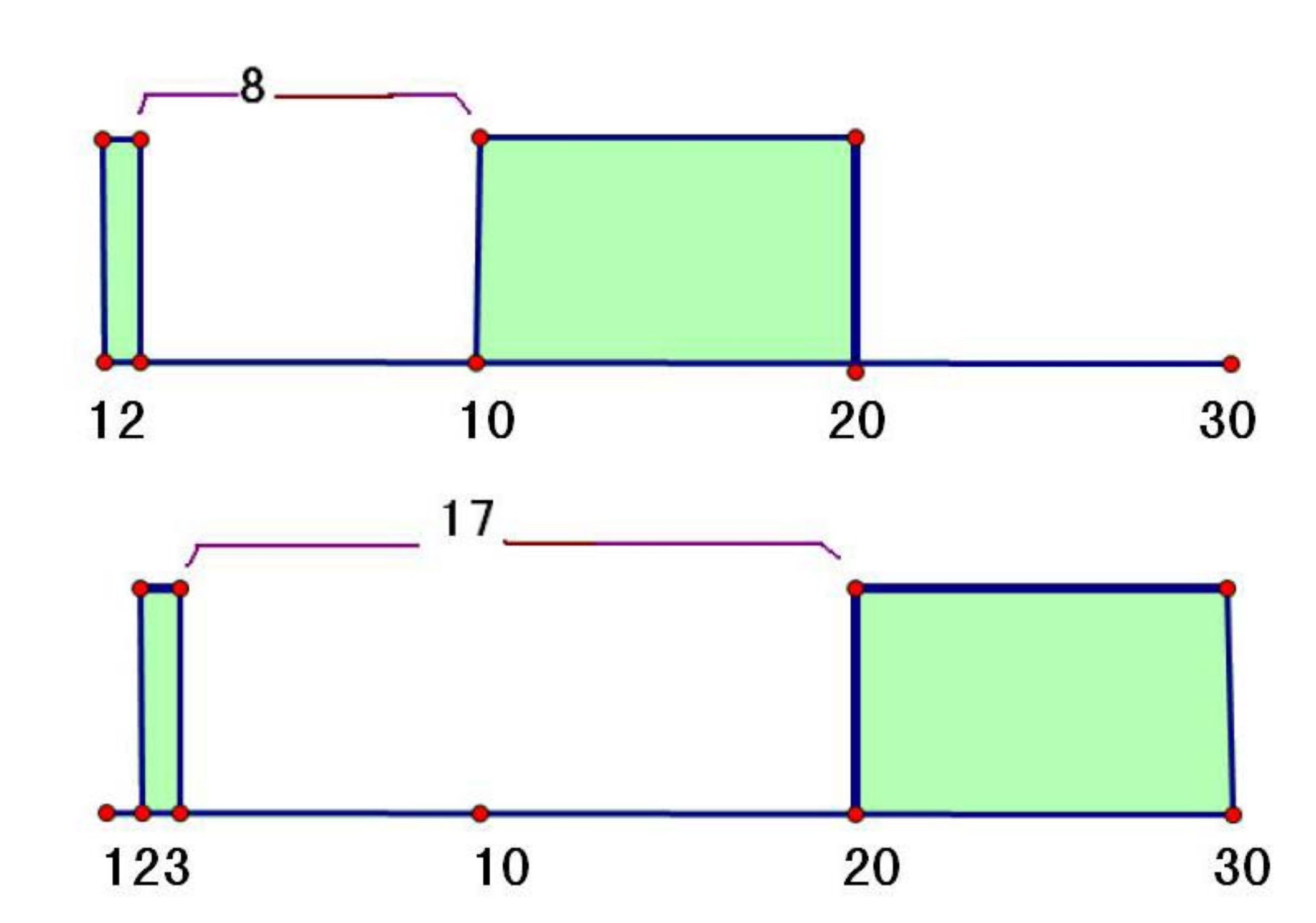}
\caption{Images of $g_1(x)$ (upper) and $g_2(x)$ (lower), from which we notice that
the gap between the colored areas in $g_2(x)$ is wider than than that is $g_1(x)$. This shows that the
distribution of $g_1(x)$ is more dense than $g_2(x)$ in the whole number range.}\label{illustration}
\end{center}
\end{figure}

Based on the above discussion, we can understand to some extent why
numbers prefer smaller first digits.  Naively one might think that the
9 digits in the decimal system play the same roles, but they define
different density $g_d(x)$ as shown above, thus behave differently
in the decimal system. For better illustration, we draw the images
of $g_1(x)$ and $g_2(x)$ in the interval $[1,30)$, as shown in Fig.~\ref{illustration}, from which we notice that the two density
functions have totally different shapes. Neither of them can simply
be a translation or an expansion of the other.

All the above derivations are rigorous. In fact, using
Eq.~(\ref{sum}) or Eq.~(\ref{sumDisForm}), we can calculate $P_d$
for any given $F(x)$ numerically. Usually, it does not strictly fit
in with Eq.~(\ref{benford}). In this sense, Benford's law is not a
rigorous ``law" with strong predictive power. However,
%In the following sections,
by using the technique of Laplace
transform, we show in the following that Benford's law is a rather good
approximation for those well-behaved probability density functions.

%\section*{Derivation for Well-Behaved Distributions}

%In this section,
We now prove that if a probability density
function has an inverse Laplace transform, it satisfies Benford's
law well. Recalling the complex inversion formula~\cite{bca}, if
F(x), extended to the complex plane, satisfies:
\begin{enumerate}
\item F(x) is analytic on $\mathbb{C}$ except for a finite number of isolated singularities;
\item F(x) is analytic on the half plane $\{x|\mathrm{Re} z > 0 \}$;
\item There are positive constants $M$, $R$, and $\beta$ such that $|F(x)|\le M/|x|^{\beta}$
whenever $|z|\geq R $,
\end{enumerate}
F(x) has an inverse Laplace transform.

We call a probability density function ``well-behaved" if it
satisfies these three conditions and its inverse Laplace transform is smooth enough, i.e.,
without violent oscillation.
Exponential functions, some fractional functions, and a handful of other common functions are
all well-behaved. Thus, the derivation in the following has wide application.
In what follows, we assume that $F(x)$ is well-behaved.

Let $f(t)$ be the inverse Laplace transform of $F(x)$, and
$G(t)$ be the Laplace transform of $g(x)$, i.e.,
\begin{equation}\label{tansDef}
F(x)=\int_{0}^{\infty} f(t)e^{-tx} {\rm d} t\,, \\
\end{equation}
\begin{equation}\label{transDef}
G(t)=\int_{0}^{\infty} g(x)e^{-tx} {\rm d} x\,.\\
\end{equation}
Laplace transforms have the following property
\begin{eqnarray}\label{tansChange}
\int_{0}^{\infty} F(x)g(x) {\rm d} x & =& \int_{0}^{\infty}{\rm d} x
g(x) \int_{0}^{\infty} f(t)e^{-tx} {\rm
d} t \nonumber\\
&=&\int_{0}^{\infty}{\rm d} t f(t) \int_{0}^{\infty} g(x)e^{-tx}
{\rm
d} x \nonumber \\
&=& \int_{0}^{\infty} f(t)G(t) {\rm d} t\,,
\end{eqnarray}
which means that Laplace transform can act on either the
function $f$ or $g$ with the above integral result keeping unchanged.

To derive the left-hand side of the above equation, we
would like to calculate the right-hand side instead. Because
it is comparably convenient to calculate the Laplace transform
of $g_d(x)$,
\begin{eqnarray}\label{gCal}
G_d(t)&=&\int_{0}^{\infty} g_d(x)e^{-tx} {\rm d} x \nonumber\\
&= &\sum_{n = -\infty}^{\infty} \int_{d \cdot 10^n}^{(d+1) \cdot
10^n}
e^{-tx} {\rm d} x \nonumber\\
&=&\frac{1}{t}\sum_{n = -\infty}^{\infty}(e^{-td \cdot 10^n} -
e^{-t(d+1) \cdot 10^n})\, ,
\end{eqnarray}
%$G_d(t)$
which can be treated as a function of two variables $d$ and $t$.
Although $d$ is defined on the decimal digit set ${1,2,...,9}$, it can
be extended to the whole real axis. Therefore, $G_d(t)$ is a continuous
function of $d$ as well as $t$. A technique to evaluate $G_d(t)$ is to
 calculate its partial derivative with respect to $d$ approximately, and then
 integrate the partial derivative to derive the result
\begin{eqnarray}\label{gPartial}
\frac{\partial G_d(t)}{\partial d} &=& \sum_{n = -\infty}^{\infty}(-
10^n e^{-td \cdot 10^n} +
 10^n e^{-t(d+1) \cdot 10^n}) \nonumber\\
&\simeq& \int_{ -\infty}^{\infty}(- 10^x e^{-td \cdot 10^x}
+ 10^x e^{-t(d+1) \cdot 10^x}) {\rm d} x \nonumber\\
&=& \frac{1}{\rm {ln}10}\int_{0}^{\infty}(-e^{-tdy} +
e^{-t(d+1)y}) {\rm d} y \nonumber \\
&=& \frac{1}{\rm{ln}10}(-\frac{1}{td}+\frac{1}{t(d+1)}).
\end{eqnarray}
%Here,
There is one and the only one approximation, i.e., we adopt an
integration to replace a summation. Because $G_d(t)\to 0$ when $d\to
\infty$, Eq.~(\ref{gPartial}) can be integrated to yield
\begin{equation}\label{gInt}
G_d(t)\simeq \frac{1}{t} \log_{10}(1+\frac{1}{d}) .
\end{equation}

Then using Eq.~(\ref{tansChange}), we obtain
\begin{eqnarray}
P_d& =& \int_{0}^{\infty} F(x)g_d(x) {\rm d} x \nonumber \\
&=&\int_{0}^{\infty} G_d(t)f(t) {\rm d} t \nonumber\\
& \simeq & \int_{0}^{\infty} \frac{f(t)}{t} \log_{10}(1+\frac{1}{d}) {\rm d} t \nonumber\\
&=&  \log_{10}(1+\frac{1}{d}) \int_{0}^{\infty} \frac{f(t)}{t} {\rm
d}t \nonumber\\
&=&  \log_{10}(1+\frac{1}{d})\,,
\label{pCal}
\end{eqnarray}
where we have used the following normalization condition of $f(t)$,
\begin{eqnarray}\label{fNorm}
1 &=& \int_{0}^{\infty} F(x) {\rm d} x \nonumber\\
&=& \int_{0}^{\infty}  {\rm d}x\int_{0}^{\infty} f(t)e^{-tx} {\rm d} t \nonumber\\
&=& \int_{0}^{\infty}  {\rm d}t f(t)\int_{0}^{\infty} e^{-tx} {\rm d} x \nonumber\\
&=& \int_{0}^{\infty} \frac{f(t)}{t} {\rm d} t .
\end{eqnarray}

Eq.~(\ref{pCal}) is exactly the first digit law for the decimal
system. Thus we show that well-behaved functions satisfy Benford's
law approximately. A more rigorous derivation without the approximately
equal signs in Eqs.~(\ref{gPartial}), (\ref{gInt}), (\ref{pCal}) can be found in Ref.~\cite{Cong19PLA}.

%{\color{blue}
Compared to Ref.~\cite{Cong19PLA}, the method provided above accords with
our intuition better. In fact, unnecessary complicated treatments
are introduced to guarantee the strictness of the proof in Ref.~\cite{Cong19PLA}.
For example, a logarithmic scale is adopted after Laplace transform, merely to
derive Eq.~(12) of Ref.~\cite{Cong19PLA}, which corresponds to Eq.~(\ref{pCal}) in this paper.
Eq.~(\ref{pCal}), though approximately holds, is set up on the original linear scale,
thus manifests itself as a property of the direct Laplace transform, instead of the logarithmic
Laplace transform which bears less intuitive physical meanings.
In this paper, no logarithmic transform is required to derive Benford¡¯s law.

According to derivations so far, we can already explain the rationality of Benford's
law through a clear chain of logic, as follows:
\begin{enumerate}
    \item The integral of the product of $F(x)$ and $g(x)$ equals
    the integral of the product of the inverse Laplace transform of $F(x)$
    and the Laplace transform of $g(x)$, i.e., Eq.~(\ref{tansChange}).
    \item The Laplace transform of $g(x)$ approximately equals the Benford term
    divided by $t$, i.e., Eq.~({\ref{gInt}}).
    \item The normalization condition of $F(x)$ guarantees that
    the integral of the inverse Laplace transform of $F(x)$ divided by $t$ equals $1$, i.e., Eq.~({\ref{fNorm}}).
    \item Therefore, the integral of the product of $F(x)$ and $g(x)$
    approximately equals the Benford term, i.e., Eq.~({\ref{pCal}}).
\end{enumerate}
Such a chain of logic is not apparent in Ref.~\cite{Cong19PLA}.
%}

%\section*{Generalization}

The second significant digit law was also given by
Newcomb~\cite{n81}. In the decimal system, it is
\begin{widetext}
\begin{equation}\label{second}
P(\mathrm{2nd~digit~being~}d) = \sum_{k=1}^9
\log_{10}(1+(10k+d)^{-1})\,, d=0,1,...,9.
\end{equation}
\end{widetext}
Hill derived a general $i$th-significant digit law \cite{h95c}:
letting $D_i$ ($D_1,D_2,...$) denote the $i$th-significant digit
(with base 10) of a number (e.g. $D_1(0.0314) = 3$, $D_2(0.0314)=1$,
$D_3(0.0314) = 4$), then for all positive integers $k$ and all $d_j
\in {0,1,...,9}$, $j=1,2,...k$, one has
\begin{equation}\label{hill}
P(D_1=d_1,...,D_k=d_k)=\log_{10}[1+(\sum_{i=1}^k d_i \cdot
10^{k-i})^{-1}].
\end{equation}

%In this section,
We propose here a general form of digit law, and show
that both the second significant digit law and the general
$i$th-significant digit law are only corollaries of this general
form.

We calculate $P_{b,d,l,k}$, which is the probability that
the integer composed of the first $k$ digits (base $b$) of an arbitrary number
[e.g. for the number $0.0314$ and $k=2$, this integer is $31$] is between $d$ and $d+l$
($b^{k-1} \le d< d+l < b^k$). Correspondingly we introduce the density function $g_{b,d,l,k}(x)$ as
\begin{equation}\label{gDefineGeneral}
g_{b,d,l,k}(x)=\sum_{n = -\infty}^{\infty}[\eta (x-d \cdot b^n)-
\eta (x-(d+l) \cdot b^n)] \,,
\end{equation}
where the right hand side is independent of $k$ (while $k$
puts restrictions on $d$ and $l$). Thus we can omit the subscript $k$ in the following.

Similar technique gives the Laplace transform of $g_{b,d,l}(x)$
\begin{equation}\label{gCalGeneral}
G_{b,d,l}(t)\simeq \frac{1}{t} \log_{b}(1+\frac{l}{d}).
\end{equation}
Thus we arrive at the general significant digit law
\begin{equation}\label{pCalGeneral}
P_{b,d,l,k}=  \log_{b}(1+\frac{l}{d})\,.
\end{equation}

We find that Benford's law (\ref{benbase}) corresponds to a special
case of this general form for $k=1$ and $l=1$, whereas Hill's
general $i$th-significant law (Eq.~(\ref{hill})) corresponds to the
case for $b=10$, $d=\sum_{i=1}^k d_i \cdot 10^{k-i}$ and $l = 1$.
Newcomb's second significant digit law can be considered as a
corollary of Hill's law according to the addition principle in
probability theory.

%\section*{Error Estimation}

%Of course, one can produce a set of telephone numbers with artificial choice to meet Benford's law.

We now calculate the error brought by our replacement of the
summation to the integration in Eq.~(\ref{gPartial}). Since
Eq.~(\ref{sumDisForm}) is always an accurate expression, the total
error is
\begin{equation}\label{totalError}
\Delta_{\mathrm{total}}=\int_{0}^{\infty} F(x)g_{b,d,l}(x) {\rm d} x
-\log_{b}(1+\frac{l}{d}).
\end{equation}
If we define
\begin{equation}\label{deltaDefine}
\Delta_{b,d,l}(t)=tG_{b,d,l}(t)-\log_{b}(1+\frac{l}{d})\,,
\end{equation}
the total error can be written as
\begin{eqnarray}\label{errorCal1}
\Delta_{\mathrm{total}} &=& \int_{0}^{\infty}\frac{ f(t)}{t}[t G_{b,d,l}(t)
-\log_{b}(1+\frac{l}{d})]{\rm d} t \nonumber\\
&=& \int_{0}^{\infty}\frac{ f(t)}{t}\Delta_{b,d,l}(t){\rm d} t.
\end{eqnarray}

Checking the definitions of the two terms of $\Delta_{b,d,l}$, we find that
the variables of both of them can be multiplied by $b$ and the
results keep unchanged, i.e., $\Delta_{b,d,l}$ is scale invariant. Hence
\begin{equation}\label{deltaPeriod}
\Delta_{b,d,l}(bt)=\Delta_{b,d,l}(t)\,.
\end{equation}

For clarity, we define
\begin{eqnarray}
\label{tToS}
&t=e^s,\\
&\widetilde{\Delta}_{b,d,l}(s)=\Delta_{b,d,l}(e^s),\\
&\widetilde{f}(s)=f(e^s).
\end{eqnarray}
The corresponding normalization condition is
\begin{eqnarray}\label{tToS2}
\int_{-\infty}^{+\infty}\widetilde{f}(s) {\rm d} s=1,
\end{eqnarray}
and the property Eq.~(\ref{deltaPeriod}) becomes
\begin{equation}\label{deltaTildePeriod}
\widetilde{\Delta}_{b,d,l}(s+\ln
{b})=\widetilde{\Delta}_{b,d,l}(s)\,.
\end{equation}
Clearly, $\widetilde{\Delta}_{b,d,l}$ is a function of period $\ln
b$. Furthermore, according to the result for exponential distribution
 in Ref.~\cite{el03} (Corollary 2, $\widetilde{f}(s)$ here is exactly
 $h_1(x)$ in Ref.~\cite{el03}, $|\widetilde{\Delta}_{10,d,1}(s)|$
 is $|h_1(x)-\log_{10}(1+\frac{1}{d})|$ in the equation of
 Corollary 2), a rather good
estimation can be made when $b=10$ and $d= l = 1$
\begin{equation}\label{deltaTildePeriod1}
 0.029< \max|\widetilde{\Delta}_{10,1,1}(s) |<0.03.
\end{equation}

We notice that the
total error can be expressed as
\begin{eqnarray}\label{errorCal2}
\Delta_{\mathrm{total}} =
\int_{-\infty}^{+\infty}\widetilde{\Delta}_{b,d,l}(s)\widetilde{f}(s){\rm
d}s,
\end{eqnarray}
where $\widetilde{f}(s)$ is dependent on $F(x)$ ultimately. In most
cases, the correlation between $\widetilde{f}(s)$ and
$\widetilde{\Delta}_{b,d,l}(s)$ is small, so is the total error.
Therefore, Benford's law can be a rather good approximation.
However, if $\widetilde{f}(s)$ is close to a periodic function with
the exact period $\ln b$, or $\widetilde{f}(s)$ changes signs very fast
between positive and negative numbers (this may happen when $F(x)$ is
artificially chosen, as the case of telephone numbers in a given
region), the small $\widetilde{\Delta}_{b,d,l}(s)$ is counted and accumulated
for many times, therefore the correlation becomes large.
%{\color{blue}
Similar problems also exist for some special types of probability density functions, whose
inverse Laplace transforms oscillate violently between positive and negative
numbers, e.g., uniform distributions or normal distributions with small variances.
Number sets drawn from such distributions, e.g., heights or ages of people,
though being natural, still violate Benford's law. %}
By arguing this, we
point out that although the above derivation seems quite general, it cannot be universally true.
More rigorous discussions about the err term with general applications to four types of number sets
can be found in Ref.~\cite{Cong19PLA}.

A special case is when the integral of $\widetilde f (s) $ is not
only convergent to 1, but also absolutely convergent to a positive
real number $M$, then
\begin{eqnarray}\label{errorCal3}
\Delta_{\mathrm{total}} &\le&
\int_{-\infty}^{+\infty}|\widetilde{\Delta}_{10,1,1}(s)||\widetilde{f}(s)|{\rm
d} s \nonumber\\
&\le & \int_{-\infty}^{+\infty}0.03|\widetilde{f}(s)|{\rm
d} s \nonumber\\
&=& 0.03M.
\end{eqnarray}
If $f(s)$ is a positive or negative definite function, it is
absolutely integrable. Such an $F(x)$ is called the completely monotonic function in mathematics.
This means that $M$ is 1, thus
$\Delta_{\mathrm{total}}$ is not greater than 0.03. Consequently
Benford's law is a good estimation. For example, when
$F(x)=\frac{2}{\sqrt{x}}e^{-\sqrt{x}}$,
$f(t)=\frac{2}{\sqrt{\pi t}}e^{-\frac{1}{4t}} >0$, and
when $F(x)=\frac{4}{(x+1)^5}$, $f(x)=\frac{e^{-t}}{6}>0$. We can assert
that in these cases, the total errors are less than 0.03. In fact, numerical
results are 0.0005 and 0.009. This verifies our estimation.

%{\color{blue}
As a rule of thumb, distributions with monotonic decreasing and relatively smooth probability density functions
often conform to Benford's law well~\cite{Cong19PLA}. Inverse Laplace transforms of such probability density functions generally
change signs only for finite times, thus being absolutely convergent.
To understand this, one can view inverse Laplace transform as
decomposing the original probability density function into a series of exponential functions, among which
some are positive and others negative. If a monotonic decreasing probability density function is
relatively smooth, i.e., without a sharp change of probability density, it can be
approached mainly by positive exponential distributions, therefore its inverse
Laplace transform does not oscillate between positive and negative numbers very much.
As an application of this rule of thumb, for non-monotonic
decreasing distributions, we can transform them into
monotonic decreasing distributions, resulting in better performance of Benford's law,
e.g., for normal distributions, we can subtract the mean value from the original data set
and obtain a monotonic decreasing distribution.
%}

The above calculations and derivations tell us that the significant
digit behaviors demonstrate that although our nature has no
preference to any specific number, it does have discrimination to
digits in numbers as a logical consequence of human's counting
system. Therefore our results justify the conventional wisdom that
the violation of Benford's law is a sign that a table of numbers is
artificial or anomalous. The underlying reason for the uneven
distribution of first digits is due to the basic property of digital
system, but not some dynamic source behind the nature as people
suspected. This also explains why we can use Benford's law to
distinguish anomalies or unnaturalness in artificial numbers.

%{\color{blue}
The mathematical expressions and derivations provided in this paper are simple, elegant, and all with clear
intuitive pictures. They are easily comprehensible.
% even by those who first learn \emph{methods of mathematical physics}.
Therefore, this version of proof of Benford's law can also
serve as an example for the application of the Laplace transform.
%}

%\section*{Summary}

The first digit law reveals an astonishing regularity in
realistic numbers. We provide in this Letter a proof of
this law from the Laplace transform, and point out the condition for the validity of the law.
Compared to Ref.~\cite{Cong19PLA}, the derivation in this Letter is simple and elegant, and
it directly reveals the rationality of the first digit law.
From our work, the first digit law is due to the basic structure of
the number system. Thus the first digit law is a general rule that
applies to vast data sets in natural world as well as in human
social activities. It is not strange anymore why
%people has found so
%many data sets in various domains to satisfy Benford's law almost
%perfectly
Benford's law is so successful in various domains.
% of human knowledge.
Such a law should be regarded as a basic
mathematical knowledge with great potential for vast applications.

%\section*{Acknowledgments}

%\begin{acknowledgments}

%This work is supported by National Natural Science Foundation of
%China (Nos.~11021092, 10975003, and 11035003), and National Fund for
%Fostering Talents of Basic Science (Nos.~J1030310 and J0730316).

%\end{acknowledgments}

\bibliography{99}

\end{document}